\documentclass[]{article}

\usepackage[nohead,margin=1.0in]{geometry}
\usepackage[utf8]{inputenc} 
\usepackage{tikz}
\usetikzlibrary{decorations.pathreplacing,positioning,arrows,backgrounds}
\usepackage{siunitx}
\usepackage{pgfplots}
\usepackage{lmodern}
\usepackage{booktabs}
\usepackage[ruled, vlined,linesnumbered]{algorithm2e}
\usepackage{setspace}
\usepackage[hidelinks]{hyperref}
\usepackage{subcaption}
\usepackage{amsthm,amsfonts,amsmath}
\newtheorem{thm}{Theorem}
\usepackage[square,numbers,sort&compress]{natbib}

\def\RR{\mathbb{R}}
\def\ZZ{\mathbb{Z}}

\newcommand{\norm}[1]{\left\lVert#1\right\rVert}
\newcommand{\ip}[2]{\left\langle#1,#2\right\rangle}
\newcommand{\abs}[1]{\left| #1 \right|}

\DeclareMathOperator*{\argmin}{arg\,min}

\newcommand{\Ltwo}{L^2(\RR)}
\newcommand{\ts}{a}   
\newcommand{\tsi}{k}  
\newcommand{\fs}{b}   
\newcommand{\fsi}{l}  
\newcommand{\tii}{{\ts\tsi}}
\newcommand{\fii}{{\fs\fsi}}
\newcommand{\kl}{{\tsi,\fsi}}
\renewcommand{\L}{L}

\newcommand{\ds}{d}
\newcommand{\dii}{{\ds^\fsi}}
\newcommand{\tfop}{\Phi}

\renewcommand{\cite}{\citep}

\newcommand{\tikzplot}[2]{ \begin{tikzpicture}
\begin{axis}[%
width=1.5in,
height=1.5in,
at={(0in,0in)},
scale only axis,
axis on top,
separate axis lines,
every outer x axis line/.append style={black},
every x tick label/.append style={font=\color{black}},
every x tick/.append style={black},
xmin=0.5,
xmax=30.5,
every outer y axis line/.append style={black},
every y tick label/.append style={font=\color{black}},
every y tick/.append style={black},
y dir=reverse,
ymin=0.5,
ymax=30.5,
axis line style={draw=none},
xlabel={#2},
xlabel style={yshift=4mm},
ticks=none,
legend style={legend cell align=left, align=left, draw=black}
]
\addplot [forget plot] graphics [xmin=0.5, xmax=30.5, ymin=0.5, ymax=30.5] {#1};
\end{axis}
\end{tikzpicture}
}

\begin{document}

\title{Peak detection for MALDI mass spectrometry imaging data using sparse frame multipliers}

\author{Florian Lieb\thanks{Department of Engineering \& Technomathematics, TH Aschaffenburg, 63743 Aschaffenburg, Germany. Corresponding author: \texttt{florian.lieb@th-ab.de}} , Tobias Boskamp\thanks{Center for Industrial Mathematics, University of Bremen, 28359 Bremen, Germany} ~and Hans-Georg Stark$^*$}

\maketitle

\begin{abstract}
  MALDI mass spectrometry imaging (MALDI MSI) is a spatially resolved analytical tool for biological tissue analysis by measuring mass-to-charge ratios of ionized molecules. With increasing spatial and mass resolution of MALDI MSI data, appropriate data analysis and interpretation is getting more and more challenging. A reliable separation of important peaks from noise (aka peak detection) is a prerequisite for many subsequent processing steps and should be as accurate as possible. 
We propose a novel peak detection algorithm based on sparse frame multipliers, which can be applied to raw MALDI MSI data without prior preprocessing. The accuracy is evaluated on a simulated data set in comparison with a state-of-the-art algorithm. These results also show the proposed method's robustness to baseline and noise effects. In addition, the method is evaluated on two real MALDI-TOF data sets, whereby spatial information can be included in the peak picking process. 
\end{abstract}

\section{Introduction}

Matrix assisted laser desorption/ ionization time-of-flight mass spectrometry imaging (MALDI-TOF MSI) is widely used for molecular imaging in drug development, discovery of medical biomarkers and histopathological analysis of tissue \cite{Alexandrov2013a,Alexandrov2009}. A two-dimensional spatially resolved molecular analysis is typically based on a pixel by pixel collection of individual spectra. By serial sectioning of sample tissue and a 2D analysis of each section, even three-dimensional analysis is possible \cite{Oetjen2015}. This leads to huge data amounts: The number of pixels (spots) may ranges from $10^4$ in the 2D case to $10^6$ for 3D, with each spectrum containing $10^3$ - $10^6$ mass-to-charge ($m/z$) bins \cite{ Wijetunge2015}. With increasing spatial and $m/z$-resolution, efficient processing and preprocessing of MALDI MSI data is highly challenging, but also inevitable. 

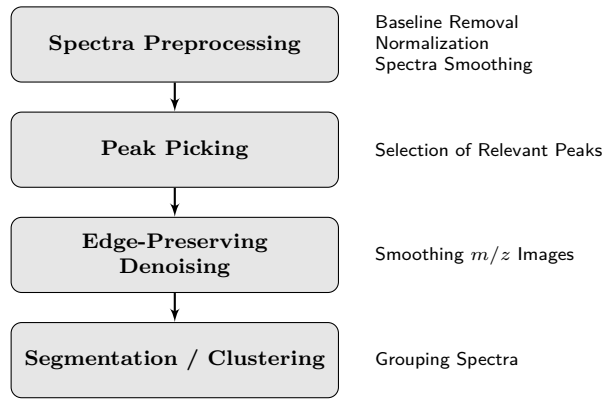
\begin{figure}[t]
	\centering
	\begin{tikzpicture}
		[node distance = 0.7cm, auto,font=\footnotesize, 
		every node/.style={node distance=1.4cm},
		comment/.style={rectangle, inner sep= 5pt, text width=4cm, node distance=0.2cm, font=\scriptsize\sffamily},
		force/.style={rectangle, rounded corners, draw, fill=black!10, inner sep=5pt, text width=4.cm, text
		badly centered, minimum height=1.cm, font=\bfseries\footnotesize\rmfamily}] 
		\node [force] (p1)              {Spectra Preprocessing};
		\node [force,below of= p1] (p3) {Peak Picking};
		\node [force,below of= p3] (p5) {Edge-Preserving Denoising};
		\node [force,below of= p5] (p6) {Segmentation / Clustering};
		\node [comment,right=3mm of p1 ] (c1) {Baseline Removal\\ Normalization \\ Spectra Smoothing};
		\node [comment,right=3mm of p3 ] (c2) {Selection of Relevant Peaks};
		\node [comment,right=3mm of p5 ] (c3) {Smoothing $m/z$ Images};
		\node [comment,right=3mm of p6 ] (c4) {Grouping Spectra};
		\path[->,thick,>=latex']
		(p1) edge (p3)
		(p3) edge (p5)
		(p5) edge (p6);
	\end{tikzpicture}
	\caption{Typical MALDI MSI processing pipeline (adapted from \cite{Alexandrov2010}).}%
	\label{fig:maldipipeline}%
\end{figure}

A typical MALDI MSI processing pipeline is shown in Fig. \ref{fig:maldipipeline}.
First, initial preprocessing includes baseline removal, normalization and spectral smoothing. Original raw spectra exhibit an intensity offset (baseline) for lower $m/z$ values which originates from matrix cluster fragments being formed during ionization \cite{Sun2011}. Matrix inhomogeneities are effectively reduced by normalizing each spectrum, e.g., using total ion count (TIC) normalization \cite{Deininger2011}. Spectral smoothing is typically performed by either applying wavelet based methods, Savitzky-Golay or Gaussian filters \cite{Shin2010,Sun2011}. 
After preprocessing, peak picking algorithms detect prominent peaks in the spectra and separate them from the noise background. This noise may arise from residual matrix clusters or chemical background molecules, ion suppression artifacts, or from electronic detector noise \cite{Deininger2011}. The variance of the noise is larger in lower mass regions and decreases with increasing $m/z$ values. Signal to noise is often low, making accurate peak picking quite challenging \cite{Shin2010}. Any further analysis like spatial segmentation or clustering groups of similar spectra, however, are based upon accurately detected peaks.

Many peak picking algorithms have already been suggested: Early peak picking approaches are based on simple local maxima \cite{Breen2000} or fitting Gaussian distributions to mass peaks \cite{Kempka2004}. More advanced algorithms make use of the continuous wavelet transform (CWT) \cite{Lange2006,Du2006,Antoniadis2010} or the discrete wavelet transform \cite{Coombes2005a,Kwon2008,Alexandrov2009}. In particular the CWT approach based on ridges and zero-crossings of wavelet coefficients has shown good performance \cite{Du2006,Antoniadis2010,Zhang2015}. Two independent comparisons of MALDI MSI peak picking algorithms also favor wavelet approaches \cite{Yang2009,Bauer2011}. 
A recently introduced peak detection algorithm is based on a dual-tree complex wavelet transform \cite{Wijetunge2015}. This approach has shown superior performance compared to the ridge line wavelet approach in \cite{Du2006}, the discrete wavelet transform based algorithm from \cite{Coombes2005a} and a Bayesian approach based on adaptive regression kernels \cite{House2011}. 

The peak picking algorithms described above are all based on spectra wise peak picking. Spatial information, however, potentially improves the peak picking process. Large peaks which are spatially surrounded by small peaks might be more likely to be ignored, whereas a relatively small peak in a neighborhood of larger peaks might be relevant.
Algorithms detecting peaks not spectra wise but in regard of corresponding $m/z$ images evidently improve the sensitivity compared to other spectra wise methods \cite{Alexandrov2013}. Algorithms which base the peak picking solely on spatial structures, however, have difficulties separating noisy peaks from actual tissue peaks and need further processing steps\cite{Eriksson2019}. A true hybrid peak picking algorithm which combines a spatial- and spectral-wise approach is still missing.
  
In this manuscript, we propose a novel spectra wise peak detection algorithm based on sparse approximations of frame multipliers with an option to include spatial information in the peak picking process. The spectra wise approach is shown to be robust against baseline and noise effects on the basis of simulated MALDI-TOF data. In addition, the spatially aware extension smooths $m/z$ images in an edge-preserving manner, as demonstrated on two different real MALDI-TOF data sets. This incorporates the first three processing steps in the MALDI MSI pipeline in Fig.~\ref{fig:maldipipeline} into a single step, significantly reducing computational complexity. Additionally, the parameter choice is simplified, as there are no longer separate values for baseline removal, normalization, smoothing and peak picking.

\section{Methods}
Before we can outline our peak picking algorithm, we recall some basic definitions and fix notation. In order to simplify notation we use square integrable functions in $\Ltwo$ with norm $\norm{\cdot}_2$ induced by $\ip{f}{g} = \int_\RR f(t)\overline{g(t)} dt$ for $f,g\in\Ltwo$. We define the following operators for $f\in\Ltwo$: the time shift operator $T_\tii f(t) = f(t-\tii)$, the frequency shift operator $M_\fii f(t) = e^{2\pi i t \fii} f(t)$ and the dilation operator $D_\dii(f) = \sqrt{\ds^{-\fsi}}f\left(\ds^{-\fsi}t \right)$ for $\ts,\fs \in \RR$, $\ds>1$ and $\tsi,\fsi \in\ZZ$. 
The collection $\mathcal{G}$ of all doubly indexed functions $g_{\tsi,\fsi}$ is given by $\mathcal{G}(g) = \{g_{\tsi,\fsi} \}_{\tsi,\fsi\in\ZZ}$.
Setting $g_{\tsi,\fsi} = M_\fii T_\tii g$, the resulting $\mathcal{G}(g)$ consists of time-frequency shifted versions of $g\in \Ltwo$ \cite{fest03}. On the other hand, defining $g_{\tsi,\fsi} = T_\tii D_\dii g$ results in time-scaled versions of $g\in \Ltwo$ \cite{Mallat2008}. $\mathcal{G}(g)$ is called a frame for $\Ltwo$ whenever the frame operator $S_{g,g}$
\begin{equation}
S_{g,g} f = \tfop_g^* \tfop_g f = \sum_{\tsi,\fsi\in\ZZ} \ip{f}{g_{\tsi,\fsi}}g_{\tsi,\fsi},
\label{eq:frameop}
\end{equation}
is bounded and invertible on $\Ltwo$. Note that in case of time frequency shifted atoms $g_{\tsi,\fsi}$, the analysis operator $\tfop_g$ is the Gabor transform \cite{fest03}
\begin{equation}
c_{\tsi,\fsi}^{\mbox{\tiny GAB}} = (\tfop_g^{\mbox{\tiny GAB}} f)_{\tsi,\fsi} = \ip{f}{g_{\tsi,\fsi}} = \int_\RR f(t) \overline{g(t-\tii)} e^{-2\pi i \fii t} dt,
\label{eq:gabtransform}
\end{equation} 
and in case of shifted and dilated atoms, $\tfop_g$ is the wavelet transform \cite{Mallat2008}
\begin{equation}
c_{\tsi,\fsi}^{\mbox{\tiny WAV}} = (\tfop_g^{\mbox{\tiny WAV}} f)_{\tsi,\fsi} = \int_\RR f(t) \frac{1}{\sqrt{\dii}}\overline{g\left( \frac{t-\tii}{\dii}\right)} dt,
\label{eq:wavtransform}
\end{equation} 
whenever $g$ satisfies the admissibility condition.

\subsection{Frame Multiplier}
The mathematical concept of frame multipliers are introduced in \cite{ba07}. With $g_{\tsi,\fsi}$ being chosen such that $f = \tfop_g^* \tfop_g  f$ the operator $\mathcal{M}_m: \Ltwo \rightarrow \Ltwo$ defined by
\begin{equation}
	\mathcal{M}_m f = \tfop_g^* I_m \tfop_g f = \sum\limits_{\kl\in\ZZ} m_\kl\ip{f}{g_\kl}g_\kl,
	\label{eq:framemul}
\end{equation}
is called a frame multiplier. Here, $I_m$ is a diagonal operator where the diagonal  $m_\kl  \in L^\infty\left(\RR^2\right)$ is denoted frame mask. Essentially, the frame multiplier acts in the transform domain of frame $\mathcal{G}$ by a pointwise multiplication with mask $m$ and a subsequent inverse transform. In case of Gabor frames such operators are frequently used to mask unwanted time-frequency coefficients \cite{fest03}. In a different approach, however, frame multipliers can be used to measure similarity between two signals $f_1,f_2\in\Ltwo$. Assuming that $\ip{f_1}{g_\kl}$ is non-zero and setting
\begin{equation}
f_2 = \sum\limits_{\kl\in\ZZ} m_\kl\ip{f_1}{g_\kl}g_\kl,
\label{eq:exframemul}
\end{equation}
the frame mask $m$ describes the transition from $f_1$ to $f_2$. Obviously, if $f_1 = f_2$ then $m=1$ for all $\tsi,\fsi\in\ZZ$. Otherwise, the mask $m$ indicates how much, and, in particular, at which
locations both signals differ. In general, however, coefficients $\ip{f_1}{g_\kl}$ are not necessarily non-trivial, which leads to the following regularization of the frame multiplier.
	
\subsection{Sparse Frame Multiplier Estimation}
The similarity of two signals $f_1$ and $f_2$ can be estimated by considering the following regularized minimization problem 
\begin{equation}
	\min_{m} \frac{1}{2}\norm{f_2-\mathcal{M}_m f_1}^2_2 + \lambda \norm{\abs{m} - 1}_1,
	\label{eq:geninvprob}
\end{equation}
with regularization parameter $\lambda>0$ \cite{doma14}. The regularization term $\norm{\abs{m} - 1}_1$ imposes a sparsity constraint on the mask $m$ such that $\abs{m}-1$ is sparse. This implies that for small values of $\lambda$ any difference between the transform coefficients of $f_1$ and $f_2$ is captured by the mask $m$. With increasing $\lambda$ small deviations between  transform coefficients are ignored by setting the resulting mask to 1. This way, only the most prominent differences between $f_1$ and $f_2$ lead to coefficients in $m$ with values different than 1. 

Unfortunately, the inverse problem in \eqref{eq:geninvprob} does not admit a closed form solution as the operator $\tfop^*_g$ is not injective. It is sufficient, however, to formulate this problem in the transform domain. For notational convenience let the transform coefficients of $f_1$ and $f_2$ be denoted by 
\begin{equation}
c_1 = \tfop_g f_1 = \ip{f_1}{g_\kl}, \quad\quad c_2=\tfop_g f_2 = \ip{f_2}{g_\kl},
\label{eq:tcc12}
\end{equation}
where the dependence on $\tsi$ and $\fsi$ is implicit from now on. The following theorem (a proof can be found in the appendix) derives the closed form solution of the simplified inverse problem stated in \eqref{eq:geninvprob}.
\begin{thm}
Let $c_1$ and $c_2$ be defined as in \eqref{eq:tcc12}. The simplified minimization problem of \eqref{eq:geninvprob}
\begin{equation}
m = \argmin_m \frac{1}{2} \norm{\abs{c_2}- m \abs{c_1}}^2_2 + \lambda \norm{m-1}_1,
\label{eq:invprob1}
\end{equation}
has the following closed form solution
\begin{equation}
m = \left(\frac{\abs{c_2}}{\abs{c_1}} - 1 \right)\left(1 - \frac{\lambda}{\abs{c_1}^2 \abs{\frac{\abs{c_2}}{\abs{c_1}}-1}} \right)^+ +1,
\label{eq:m}
\end{equation}
where $(\cdot)^+ = \max(0,\cdot)$ denotes the maximum with zero. 
\end{thm}

\section{Algorithm}

\subsection{Spectra-wise Approach}
The proposed peak picking algorithm can now be defined in a finite dimensional setting as follows. Let $f \in \RR^\L$ be a single raw MALDI MSI spectrum of length $\L$. This signal is divided into overlapping slices $f_i$ of length $M$ and overlap $O \in(0,1)$. An overlap between slices is required, otherwise peaks might be unintentionally separated into two consecutive slices. For both slices the transform coefficients based on the given frame $\{g_\kl\}$ are computed in the next step and the mask $m$ can be estimated using \eqref{eq:m} for a given regularization parameter $\lambda$. Now, this mask indicates the most prominent differences between the two consecutive and overlapping transform coefficients $c_i$ and $c_{i+1}$ for corresponding $f_i$ and $f_{i+1}$. If both slices are similar with respect to the chosen $\lambda$, the mask is a constant one and supposedly no peak is present. On the other hand, peaks are present whenever the mask takes values different from 1. 

In the next step, the coefficients of the mask $m$, or more precisely $m-1$, are analyzed. If no peak is present in either $f_i$ and $f_{i+1}$, the sum $z_\tsi = \sum_\fsi (m_{\tsi,\fsi} - 1)$ is zero for all $\tsi$, where $\tsi$ is associated to corresponding $m/z$ values of $f_i$. Note that $\fsi$ sums over all frequencies or scales. On the other hand, $m-1$ leads to negative values whenever a peak is present in $f_i$ and no peak in $f_{i+1}$, and positive values vice versa. If the overlap is chosen such that $O\geq0.5$, positive values of $m-1$ should give negative values in the subsequent iteration step, when estimating the mask for $c_{i+1}$ and $c_{i+2}$. This also allows to discriminate peaks whenever they are at the same relative location in $f_i$ and $f_{i+1}$.
Hence, it is sufficient to consider negative values of $m-1$ only. With $(\cdot)^{-}=\min(0,\cdot)$, computing 
\begin{equation}
	z^{i}_\tsi = \sum_\fsi \abs{\left(m^i_\kl - 1\right)^{-}},
	\label{eq:sm}
\end{equation}
for all slices of $f_i$ leads to an indicator signal of length $L$ comprising the detected peaks within the spectrum $f$.
A summary of the proposed peak picking approach can be found in Algorithm \ref{alg:maldipeakpickingalgo}. Note that the inner products in line 3 and 4 are finite dimensional with respect to the frame length $M$. Additionally, the computation for $N$ spectra can be done using a loop or, as implicitly indicated in  Algorithm \ref{alg:maldipeakpickingalgo}, a vectorized approach.

\begin{algorithm}[t]
\SetAlgoLined
\DontPrintSemicolon
\SetKwInOut{Input}{Input}
\SetKwInOut{Output}{Output}
\Input{$f \in\RR^{N\times\L} $, $\lambda \in\RR^+$,  $M$, $O$, $g_\kl$ }
\Output{$z \in \RR^{N\times\L}$}		
\BlankLine
	$K \gets $ Total number of slices of length $M$ and overlap $O$ dividing $L$\;
	$z \gets 0$\;
	\For{$i=0,1,2,\ldots,K-1$}
	{
	  $c_{1} \gets \ip{f_i}{g_\kl}$,~~~$c_{2} \gets \ip{f_{i+1}}{g_\kl}$,~~~$y \gets \abs{c_{2}}\abs{c_{1}}^{-1}$\;
		\vspace{0.1cm}
		$m_i \gets \left(y - 1 \right)\left(1 - \frac{\lambda}{\abs{c_{1}}^2 \abs{y-1}} \right)^+ +1$\;
		\vspace{0.2cm}
		$z_i \gets \max\left( z_i, \sum_l \abs{\left(m^i_{\kl} - 1\right)_{\mbox{\scriptsize neg}}} \right)$ \;
	}
\caption{MALDI Peak Picking Algorithm}
\label{alg:maldipeakpickingalgo}
\end{algorithm}

Slicing spectra into smaller parts has several advantages. First, raw spectra can be analyzed without preprocessing the baseline. If the slice length $M$ is chosen small enough, the influence of baseline effects of two consecutive slices is negligible. Additionally, the algorithm's sensitivity can be adjusted to the noise level. Based on time-frequency or time-scale coefficients for both slices, the variance of the noise in these slices can be estimated by $\tilde{\sigma} =  median\left(\left\{\ip{f_{1,2}}{g_{k,l}} \right\}_{k,l \subset \hat{\ZZ}}\right)/ 0.6745$ \cite{Mallat2008}. Here, $\hat{\ZZ} \subset\ZZ$ such that the corresponding time-frequency/time-scale representation does not contain peak information. The regularization parameter $\lambda$ can then be weighted according to this noise level. Hence, the sensitivity increases whenever the noise variance decreases within a single spectrum.

\subsection{Spatially-aware Approach}

Algorithm \ref{alg:maldipeakpickingalgo} can be modified to include information of surrounding spectra. As mentioned before, the possibility of a peak being detected is larger if the spectra of neighboring spots also contain peaks at approximately the same $m/z$ ratio, whereas a peak surrounded by noise in the spatial neighborhood might be more likely to be ignored. 

To formulate the spatial awareness mathematically, let for every spectrum $f$ the set $\mathcal{N}$ be its neighboring spectra, including the actual spectrum itself. Further, denote by $w_j$, $j\in\mathcal{N}$, a weight corresponding to each neighbor such that $\sum_{j\in\mathcal{N}} w_j = 1$. 
Defining 
\begin{equation}
\tilde{y}= \sum_{j\in\mathcal{N}} w_j \frac{\abs{c_{2,j}}}{\abs{c_{1,j}}} \quad\mbox{and}\quad \tilde{c}_1 = \sum_{j\in\mathcal{N}} w_j\  c_{1,j},
\label{eq:y_neigh}
\end{equation}
weighs coefficients $c_1$ and $c_2$ at the current spot with their neighboring coefficients $c_{1,j}$ and $c_{2,j}$. The estimation of the mask $m$ in \eqref{eq:m} can then be reformulated as
\begin{equation}
m = \left(\frac{\abs{c_2}}{\abs{c_1}} - 1 \right)\left(1 - \frac{\lambda}{\abs{\tilde{c}_1}^2 \abs{\tilde{y}-1}} \right)^+ +1.
\label{eq:m_spatial}
\end{equation}
This scales the regularization parameter $\lambda$ for each coefficient depending on the characteristics of neighboring spectra.
The weights $w$ can, for example, be a simple average kernel, where each element is defined by $w_j = \frac{1}{|\mathcal{N}|} \forall j\in\mathcal{N}$ and $|\mathcal{N}|$ denotes the cardinality of $\mathcal{N}$. Other choices of weights could include Gaussian kernels with different variances or circular average filters. Even non-linear approaches, e.g., a median filter, can be used to compute $\tilde{y}$ and $\tilde{c_1}$.

\section{Results and discussion}

\subsection{Simulated MALDI-TOF Data}

\subsubsection{Data Sets}
The performance of our proposed peak picking algorithm is evaluated based on the data set, which has been used by \cite{Wijetunge2015}. This data set is introduced in \cite{coombes2005} and is based on the physical principles of time-of-flight mass spectrometry, emulating characteristics of real MALDI-TOF data. In total, the data set consists of 2500 individual spectra with annotated peak locations each having a length of 15,000 up to 30,000 samples. Each spectrum is independent, which implies that the spatial awareness approach can not be utilized in this simulated data set. 
In order to evaluate the incorporation of spatial information in the peak picking algorithm, an entire MSI dataset has been simulated using the Cardinal toolbox \cite{Bemis2015}.

\subsubsection{Performance Measures}
In order to be consistent with the performance measures in \cite{Wijetunge2015,Yang2009}, the sensitivity is defined as 
\begin{equation}
\mbox{Sensitivity} = \frac{\mbox{Number of correctly identified peaks}}{\mbox{Number of reference peaks}}.
\label{eq:sensitivity}
\end{equation}
The False Discovery Rate (FDR) is given by
\begin{equation}
\mbox{FDR} = \frac{\mbox{Number of falsely identified peaks}}{\mbox{Number of total peaks detected}}.
\label{eq:fdr}
\end{equation}
Ideally, an optimal peak picking algorithm has sensitivity of 1 and a FDR of 0. The larger the sensitivity and at the same time the lower the FDR, the better the performance of the algorithm. Both values can be combined into a single performance measure, denoted by F1-score, defined by 
\begin{equation}
\mbox{F1-score} = \frac{2\cdot\left(1-\mbox{FDR}\right)\cdot\mbox{Sensitivity}}{1-\mbox{FDR}+\mbox{Sensitivity}}.
\label{eq:f1score}
\end{equation}
This gives a single performance value, taking the sensitivity as well as the FDR into account. In the following numerical evaluation, a peak is classified as correctly identified if it is within 1\% of the expected $m/z$ value as proposed in \cite{Wijetunge2015,Yang2009}. 

\subsubsection{Parameter Settings}
Each spectrum is divided into slices of length 60 samples with an overlap of 0.5. The frame is set to be either a Gabor or a wavelet frame. The Gabor frame is based on a Hann window with a width of 20 (60 for the  samples and a time- and frequency-sampling step size of 1 each, i.e., $\ts=1$ and $\fs=1$. Parameters for the wavelet frame are based on algorithms implemented in \cite{ltfatnote030}. The parameters $fmin$ and $bw$ are set to 1000\,Hz each and the number of bins is set to 30. The generating wavelet is the uncertainty equalizer derived in \cite{lelisost14}. The regularization parameter $\lambda$ then controls the number of detected peaks and is chosen such that the number of detected peaks equals the number of peaks inserted. 

\subsubsection{Results for the Simulated Data Sets}

The performance of our proposed algorithm using Gabor frames as well as wavelet frames based on \cite{Lieb2018} is compared to the peak picking algorithm introduced in \cite{Wijetunge2015}. Note that \citeauthor{Wijetunge2015}'s algorithm is different from the one proposed in \cite{Wijetunge2015a}. As \citeauthor{Wijetunge2015a}'s EXIMS algorithm \cite{Wijetunge2015a} only measures how well an $m/z$-image is structured, it is not appropriate to use in this context. The algorithms are applied to all 2500 spectra. The resulting mean sensitivity, FDR and F1-score can be seen in Table \ref{tab:maldires_sim}. The proposed algorithm is evaluated without and with baseline correction using the Matlab routine \texttt{msbackadj} \cite{Andrade2003}. 
\begin{table}
\caption{\label{tab:maldires_sim}Performance of the proposed peak picking algorithms in comparison with the algorithm from \cite{Wijetunge2015}. Mean values and corresponding standard errors over all 2500 spectra are shown.}
\centering
\begin{tabular}{@{}rlll}
\toprule
& Sensitivity (\%) & FDR (\%)        & F1-score (\%)  \\
\midrule
CWT \cite{Du2006} 										& $77.21\pm0.24$   & $30.84\pm0.29$  & $72.95\pm0.28$ \\
Wijetunge \cite{Wijetunge2015}  						& $86.86\pm0.18$   & $28.89\pm0.22$  & $77.29\pm0.14$ \\ 
Algorithm (Gabor)   		& $87.89\pm0.08$   & $20.70\pm0.12$  & $83.23\pm0.09$ \\ 
- with baseline removed & $94.26\pm0.07$   & $12.20\pm0.14$  & $90.76\pm0.09$ \\ 
Algorithm (Wavelet) 		& $92.13\pm0.07$   & $19.81\pm0.14$  & $85.59\pm0.10$ \\ 
- with baseline removed & $92.14\pm0.07$   & $19.65\pm0.14$  & $85.70\pm0.10$ \\
\bottomrule
\end{tabular}
\end{table}
The CWT as well as the Wijetunge results in the first two rows of Table \ref{tab:maldires_sim} reflect the ones by Wijetunge et al., see \cite[Table 3]{Wijetunge2015}. However, our approach has a higher sensitivity at a lower FDR even with no baseline removed. 
Using wavelet frames, the results in Table \ref{tab:maldires_sim} demonstrates that the baseline does not influence the accuracy of the peak picking method.
The sensitivity of the proposed algorithm yields the best performance at the lowest FDR for Gabor frames and a prior baseline removal. 

The run time of Wijetunge's algorithm is approximately 60 seconds for a single spectrum on a 2.9\,GHz i7 QuadCore processor, resulting in a computation time of more than 41 hours for all 2500 spectra. The proposed algorithm, however, takes roughly 5 minutes to process all 2500 spectra in a vectorized implementation of Algorithm \ref{alg:maldipeakpickingalgo}. In comparison, the CWT approach in \cite{Wijetunge2015} takes 14 minutes, and a simple orthogonal matching pursuit (OMP) approach from \cite{Alexandrov2010} takes just under 2 minutes in SCiLS Lab. Note that the CWT approach outperforms template based approaches like OMP as shown in \cite{Bauer2011,Yang2009}.

\begin{figure}[t]%
\centering
\tikzplot{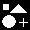}{(a)}%
\tikzplot{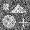}{(b)}%
\tikzplot{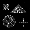}{(c)}
\tikzplot{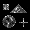}{(d)}
\caption{Performance of the spatially-aware approach based on simulated data: Ground truth with four regions corresponding to four different $m/z$ values (a), the noisy simulated data set (b), regions after peak picking (c) and after spatially-aware peak picking (d).}%
\label{fig:simspatial}%
\end{figure}

The results with a spatially structured MALDI MSI data set are visualized in Fig.~\ref{fig:simspatial}. Four regions corresponding to four different $m/z$ values have been generated (cf.~Fig.~\ref{fig:simspatial}(a)-(b)). Each of the four $m/z$ values corresponds to one shape (square, triangle, circle and cross). Note that the displayed figures are a combination of the four separate $m/z$ images. The proposed peak picking algorithm has then been applied to this dataset without and with including spatial information (using Gaussian weights $w_j$ with a  standard deviation of 0.5) leading to Fig.~\ref{fig:simspatial}(c) and Fig.~\ref{fig:simspatial}(d) respectively. While the spectra wise approach leads to regions with missing peaks, the spatially-aware approach introduces smoother regions.

\subsection{Real MALDI-TOF Data}
It is rather difficult to verify the performance of the proposed peak picking algorithm on real MALDI MSI data, since peak locations are generally not known. Nonetheless, it is possible to apply the peak picking approach to MALDI MSI data sets resulting in two possible applications: peak picking and denoising. Treating the output $z$ primarily as an indicator variable for possible peak locations leads to a classical peak picking application. Considering the output $z$, on the other hand, as actual 'spectra' itself results in a strongly denoised MALDI spectrum. It is important to note, that in this case, peak intensities in $z$ do not have any relation to peak intensities in the original spectrum and the term 'spectra' is only used to indicate the intended usage. Despite this drawback, it might be beneficial for revealing certain structures in $m/z$ images, which might be challenging to detect otherwise. Additionally, the denoised data can give an initial overview of interesting features of the data set. 

In the following we are going to consider two different MALDI MSI data sets: a linear TOF data set which has been used quite frequently in the literature and a reflector TOF data set. 

\subsubsection{Data Sets}
\paragraph{Linear TOF - Coronal Rat Brain Data Set.}
This data set is introduced in \cite{Alexandrov2010}. A 10\,\si{\micro\meter} frozen tissue section of a rat brain was prepared for MALDI MSI using sinapinic acid as a matrix. With a lateral resolution of 80\,\si{\micro\meter} 20,185 spectra were acquired, each containing 6,618 $m/z$ bins ranging from 2.5 - 25 kDa. The spectra are preprocessed by removing the baseline using a top hat filter as well as a TIC normalization before applying the proposed peak picking algorithm. No spectra smoothing has been applied. In \cite[Fig.~5C]{Alexandrov2010}, the anatomical annotation of this section is given. 

\paragraph{Reflector TOF - FFPE Lung Data Set.}
MALDI MSI based on formalin-fixed and paraffin-embedded (FFPE) tissue samples is gaining increased interest in pathological applications \cite{Aichler2015}. Sample preparation and acquisition of the following FFPE data set is similar as described in more detail in \cite{boskamp2017}. MALDI MSI data of a human lung FFPE tissue sample was obtained in positive ion reflector mode on a MALDI-TOF instrument by Bruker Daltonics (Autoflex Speed). The data set consists of 3,567 spectra, each containing 20,992 samples in the mass range of 700 - 4,000 $m/z$. The baseline has been removed prior to TIC normalization. Unfortunately, the lung data set is not annotated.

\subsubsection{Results for the Coronal Rat Brain Data Set}
The peak picking approach based on Algorithm \ref{alg:maldipeakpickingalgo} is applied to the entire rat brain data set with the following input parameters: the slice length and overlap remain as previously defined ($M = 60$  and $O =  0.5$). The Gabor frame is based on a Hann window of width 15 samples. The regularization parameter is fixed to $\lambda = 1.5\mathrm{e}{-3}$ for all spectra.
\begin{figure*}[t!]
\begin{tabular}[b]{lr}
\hspace{-1.3em}
\begin{tabular}[b]{l}
\begin{subfigure}[b]{0.47\columnwidth}
	\centering
	\includegraphics[width=\textwidth]{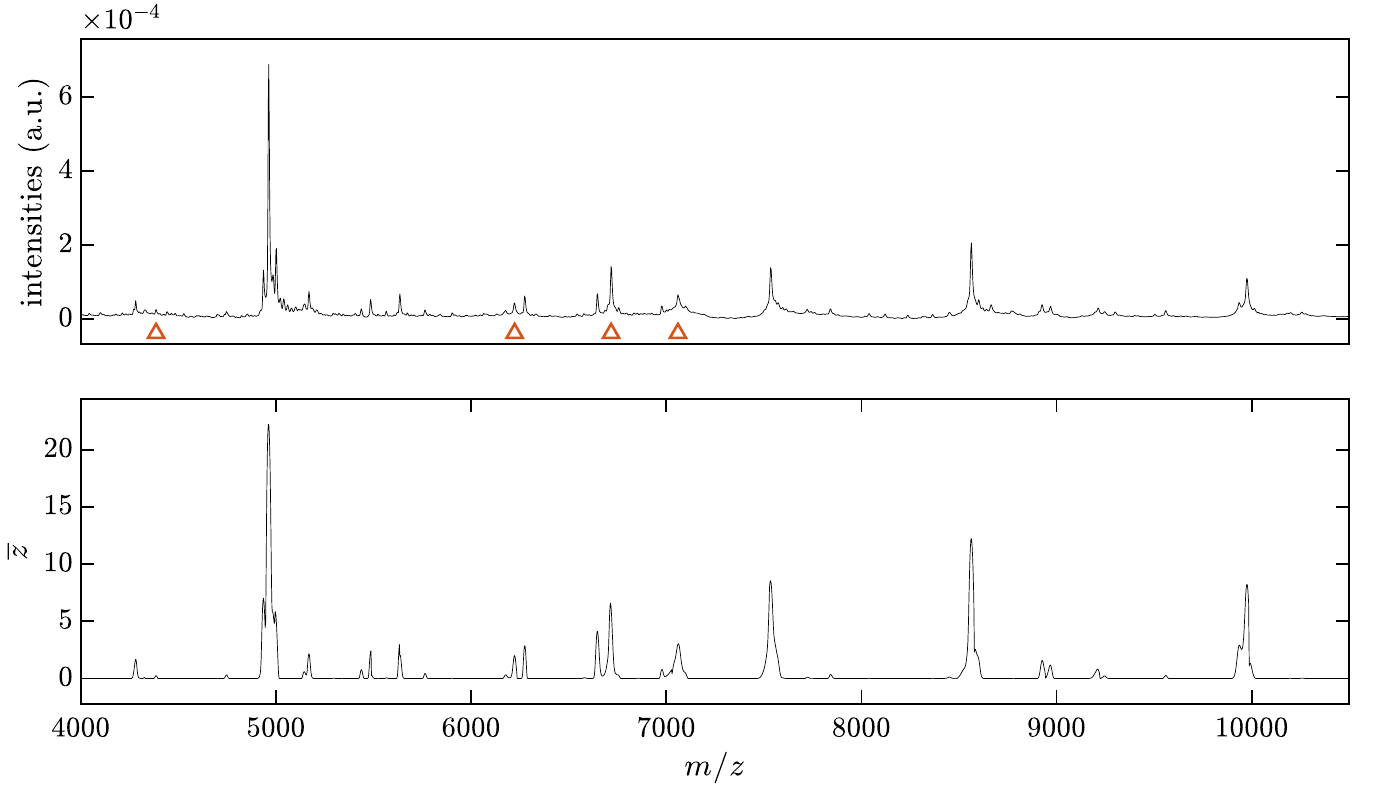}
	\caption{Mean spectrum}
 	\label{fig:A}
\end{subfigure} \\
\begin{subfigure}[b]{0.47\columnwidth}
 	\begin{tikzpicture}
 	\definecolor{fillcolor}{rgb}{0.15,0.15,0.15}
     \def\xa{-2.08}
     \def\xb{-1.92}
     \def\xc{-1.73}
     \def\xd{0.860}
     \def\xe{1.075}
     \def\xf{1.21}
     \def\xg{2.02}
     \def\xh{2.18}
		 \def\xz{-1.18}
		 \def\yz{1.5}
 	\node[inner sep=0pt](fig1) at(0,0) {\centerline{\includegraphics[width=\textwidth]{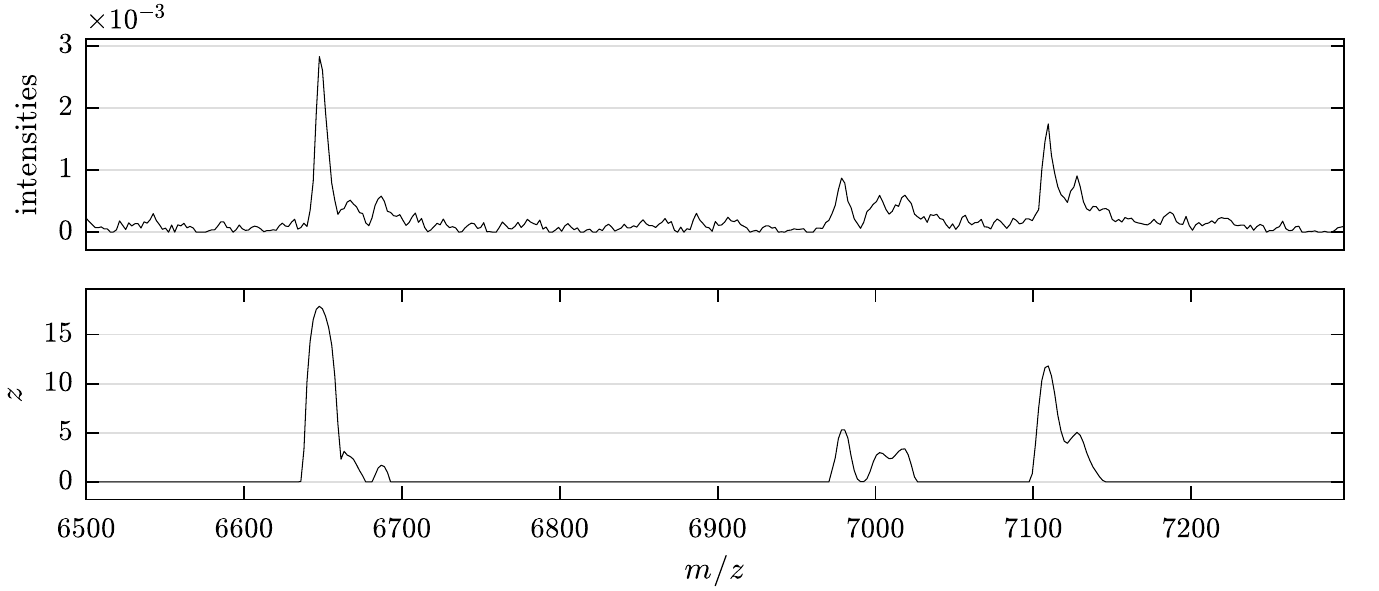}}};
	
 	\draw[color=fillcolor,dashed,opacity=0.7] (\xa,\xz)--(\xa,\yz);
 	\draw[color=fillcolor,dashed,opacity=0.7] (\xb,\xz)--(\xb,\yz);
 	\draw[color=fillcolor,dashed,opacity=0.7] (\xc,\xz)--(\xc,\yz);
	
 	\draw[color=fillcolor,dashed,opacity=0.7] (\xd,\xz)--(\xd,\yz);
 	\draw[color=fillcolor,dashed,opacity=0.7] (\xe,\xz)--(\xe,\yz);
 	\draw[color=fillcolor,dashed,opacity=0.7] (\xf,\xz)--(\xf,\yz);
	
 	\draw[color=fillcolor,dashed,opacity=0.7] (\xg,\xz)--(\xg,\yz);
 	\draw[color=fillcolor,dashed,opacity=0.7] (\xh,\xz)--(\xh,\yz);
	
 	\end{tikzpicture}
	\caption{Preservation of overlapping peak structures after peak picking}
 	\label{fig:B}
\end{subfigure}
\end{tabular}
&
\begin{subfigure}[b]{0.47\columnwidth}
	\def\fosize{\tiny}
	\def\absleft{-3.5}
	\begin{tikzpicture}
	  \node[inner sep=0pt](A) { \includegraphics[width=\textwidth]{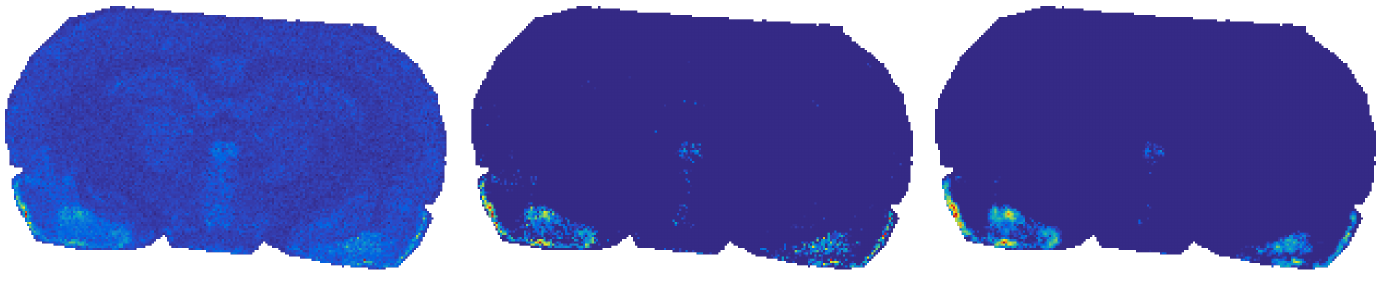} };
    \node[inner sep=0pt] at (\absleft,1) (dummy) {\fosize \textbf{(A)} $m/z$ 4386};
		\node[inner sep=0pt,yshift = 0.05cm,xshift=-2.8cm] at (A.south) {\fosize raw};
		\node[inner sep=0pt,yshift = 0.05cm,xshift=-0.1cm] at (A.south) {\fosize basic};
		\node[inner sep=0pt,yshift = 0.05cm,xshift=2.7cm] at (A.south) {\fosize average};
		\node[inner sep=0pt] at (0,-1.1) {};
	\end{tikzpicture}
		
	\begin{tikzpicture}
	  \node[inner sep=0pt](A) { \includegraphics[width=\textwidth]{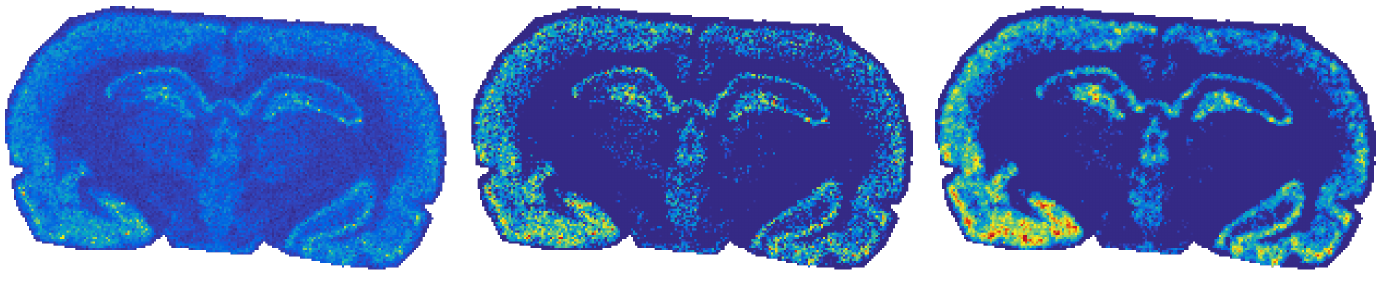} };
    \node[inner sep=0pt] at (\absleft,1) (dummy) {\fosize \textbf{(B)} $m/z$  6223};
		\node[inner sep=0pt,yshift = 0.05cm,xshift=-2.8cm] at (A.south) {\fosize raw};
		\node[inner sep=0pt,yshift = 0.05cm,xshift=-0.1cm] at (A.south) {\fosize basic};
		\node[inner sep=0pt,yshift = 0.05cm,xshift=2.7cm] at (A.south) {\fosize average};
		\node[inner sep=0pt] at (0,-1.1) {};
	\end{tikzpicture}

	\begin{tikzpicture}
		\node[inner sep=0pt](A) { \includegraphics[width=\textwidth]{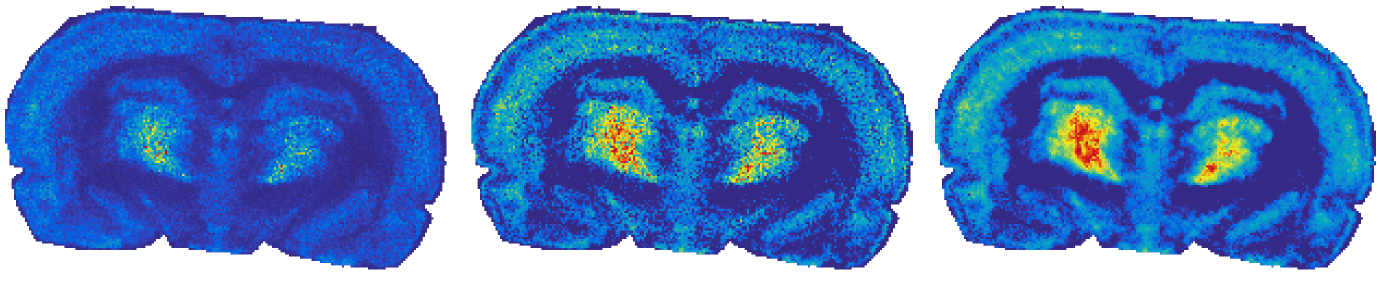} };
    \node[inner sep=0pt] at (\absleft,1) (dummy) {\fosize \textbf{(C)} $m/z$ 6717};
		\node[inner sep=0pt,yshift = 0.05cm,xshift=-2.8cm] at (A.south) {\fosize raw};
		\node[inner sep=0pt,yshift = 0.05cm,xshift=-0.1cm] at (A.south) {\fosize basic};
		\node[inner sep=0pt,yshift = 0.05cm,xshift=2.7cm] at (A.south) {\fosize average};
		\node[inner sep=0pt] at (0,-1.1) {};
		
	\end{tikzpicture}

	\begin{tikzpicture}
		\node[inner sep=0pt](D) { \includegraphics[width=\textwidth]{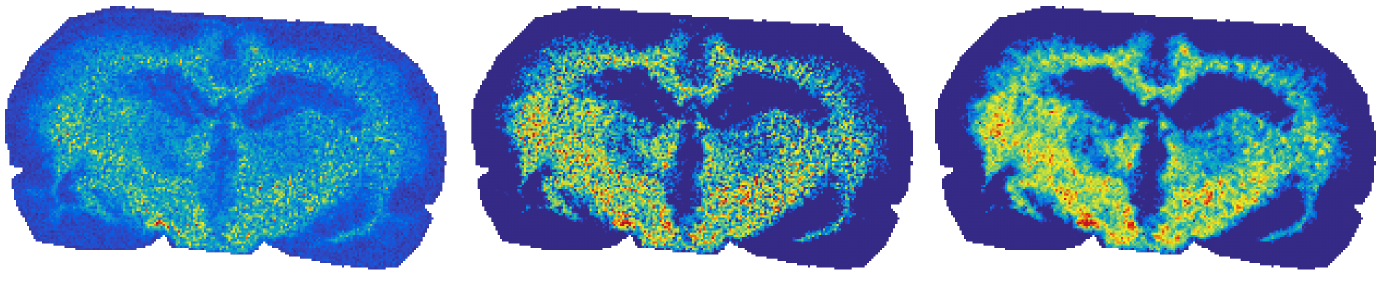} };
		\node[inner sep=0pt,yshift = 0.05cm,xshift=-2.8cm] at (A.south) {\fosize raw};
		\node[inner sep=0pt,yshift = 0.05cm,xshift=-0.1cm] at (A.south) {\fosize basic};
		\node[inner sep=0pt,yshift = 0.05cm,xshift=2.7cm] at (A.south) {\fosize average};
    
		\node[inner sep=0pt] at (\absleft,1) (dummy) {\fosize \textbf{(D)} $m/z$ 7060};
		\node[inner sep=0pt] at (0,-1.1) {};
		
	\end{tikzpicture}

	\caption{Selected $m/z$ images}
	\label{fig:C}
\end{subfigure}
\end{tabular}
\caption{Results for the coronal ratbrain data set: (a) mean spectra of the original data (top) and after peak picking (bottom), (b) an example spectrum containing overlapping peaks before and after peak picking and (c) selected $m/z$ images of the \textit{raw} data, after peak picking without spatial information (\textit{basic}) and including spatial information based on a $3\times3$ average filter (\textit{average}). Triangles indicate the selected $m/z$ values in the mean spectrum.}
\label{fig:ratbrain}
\end{figure*}
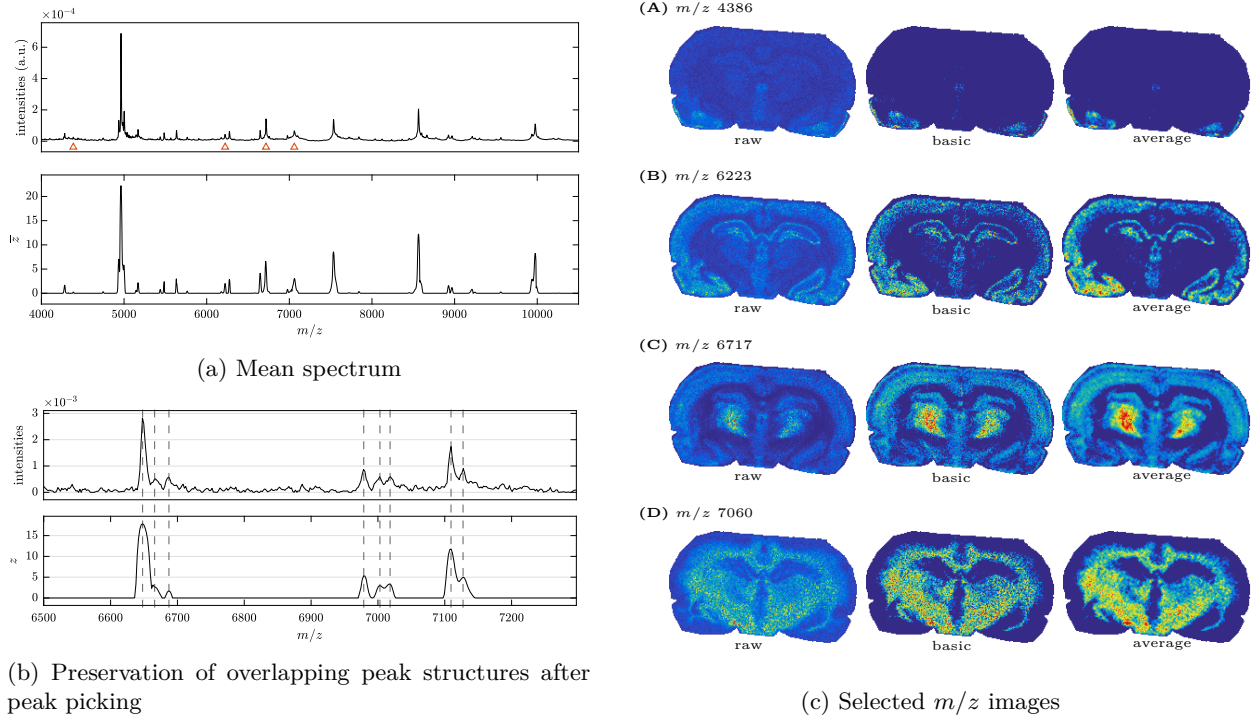
The mean spectrum $\overline{z}$ over all spots is depicted in Fig.~\ref{fig:A} (bottom) in contrast to the mean spectrum of the raw data (top). It shows the same prominent features as the original data set. However, only 48\% of all $m/z$ images contain non-trivial coefficients. This means, that for the remaining 52\% of $m/z$ values no peak is detected in any of the 20,185 spots. Clearly, this procedure is sensitive to the choice of the regularization parameter $\lambda$. Smaller values increase the sensitivity, resulting in more detected peaks per single spectrum. Larger values, on the other hand, increase the denoising effect by choosing less peaks. Recall from \eqref{eq:sm} that $z$, and hence the mean spectrum of $z$, does not reflect original intensities any more, but rather significant changes between Gabor coefficients of two consecutive slices.

In the following, the modified approach proposed in \eqref{eq:m_spatial} is based on an average filter for a $3\times3$ neighborhood $\mathcal{N}$. Hence, the filter coefficients are $w_j = \frac{1}{9}$ for all $j\in\mathcal{N}$. At the edges the neighborhood size reduces to the number of available spectra. Four selected $m/z$ images are illustrated in Figure \ref{fig:C}, showing the differences between the original data, the \textit{basic} approach based on \eqref{eq:m} and the \textit{average} approach.  Corresponding $m/z$-values are indicated with a triangle in the mean spectrum of Fig.~\ref{fig:A}. The denoising effect of the proposed algorithm is clearly visible when comparing raw data with the basic or average approach. Additionally, the neighborhood based approach smooths peak areas in $m/z$ images, while preserving edges. The sensitivity of the proposed peak picking approach is large enough to also detect low intensity peaks. Alexandrov and Bartels (\citeyear{Alexandrov2013}) showed that the low intensity peak at $m/z=4385.9$, depicted in Fig.~\ref{fig:C} (A), is not detected by other spectrum-wise peak picking approaches.

The detection of overlapping peaks is crucial whenever the sampling rate is low and isotopes are not clearly separated. Figure \ref{fig:B} shows part of the rat brain spectrum with overlapping peaks and the output of the proposed peak picking algorithm $z$. Local maxima in $z$ indicate the positions of overlapping peaks in the original spectrum.

\subsubsection{Results for the FFPE Lung Data Set}
The proposed peak picking approach is applied to the lung data set with similar parameter settings as previously used: $M=60$, $O=0.5$, a Gabor frame of length $M$ based on a Hann window of length 8 samples and a regularization parameter $\lambda=3\mathrm{e}{-3}$. 

\begin{figure*}[t!]
\setlength{\abovecaptionskip}{5pt plus 3pt minus 2pt}
\hspace{-1.5em}
\begin{tabular}[b]{cc}
\begin{tabular}[b]{c}
\begin{subfigure}[b]{0.48\columnwidth}
	\includegraphics[width=\textwidth]{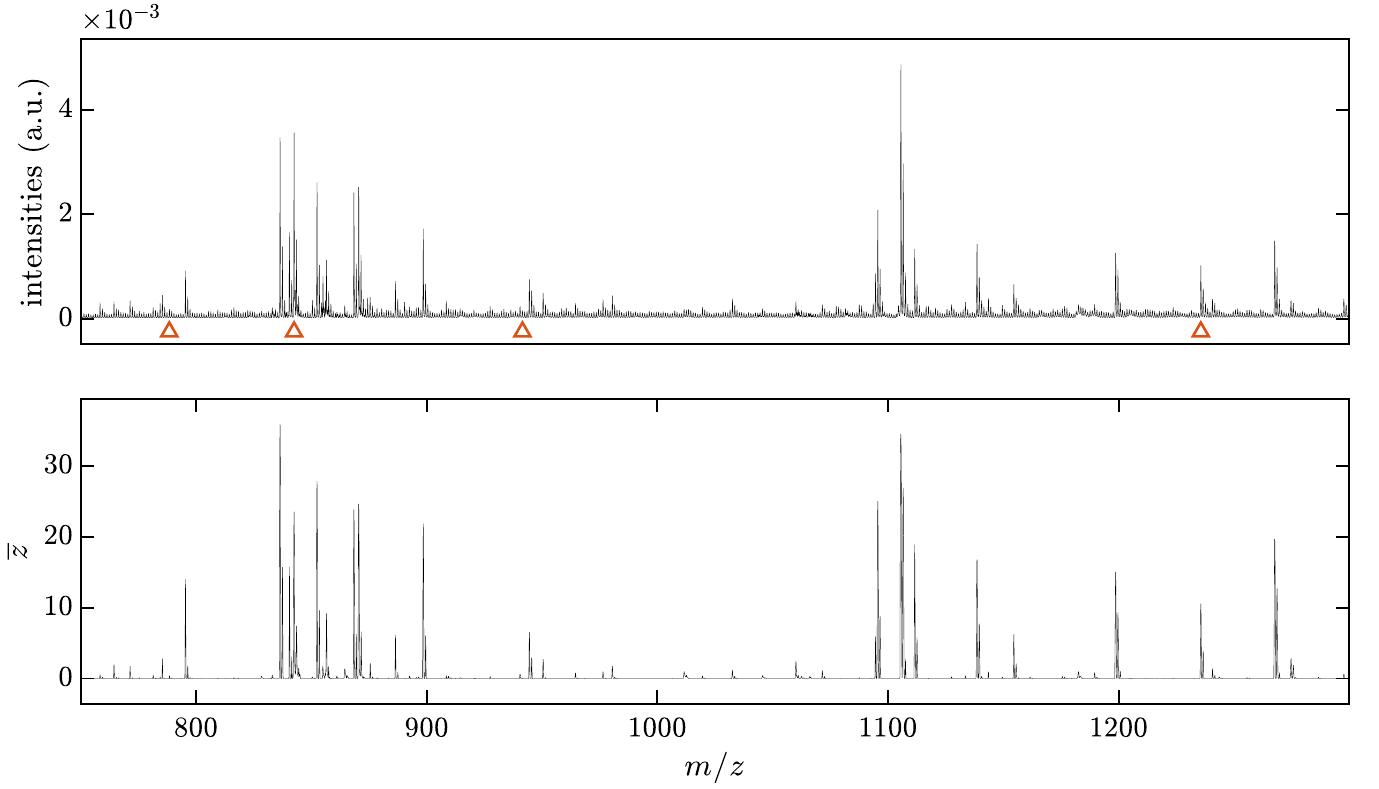}
	\caption{Mean spectrum}
	\label{fig:2A}
\end{subfigure} \\
\begin{subfigure}[b]{0.48\columnwidth}
	\includegraphics[width=\textwidth]{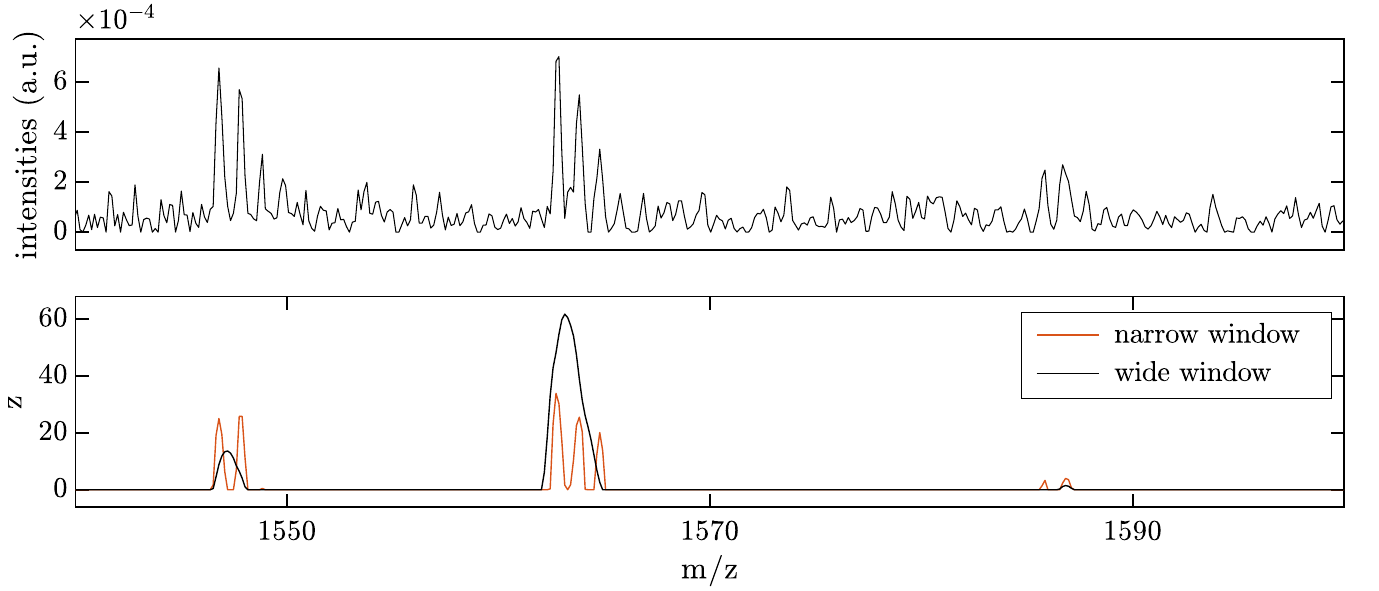}
	\caption{Detection of isotope patterns}
	\label{fig:2B}
\end{subfigure}
\end{tabular}
&
\hspace{5pt}\begin{subfigure}[b]{0.45\columnwidth}
	\def\fosize{\tiny}
	\def\absleft{-3.5}
	\begin{tikzpicture}
		\node[inner sep=0pt](A) { \includegraphics[width=0.9\textwidth]{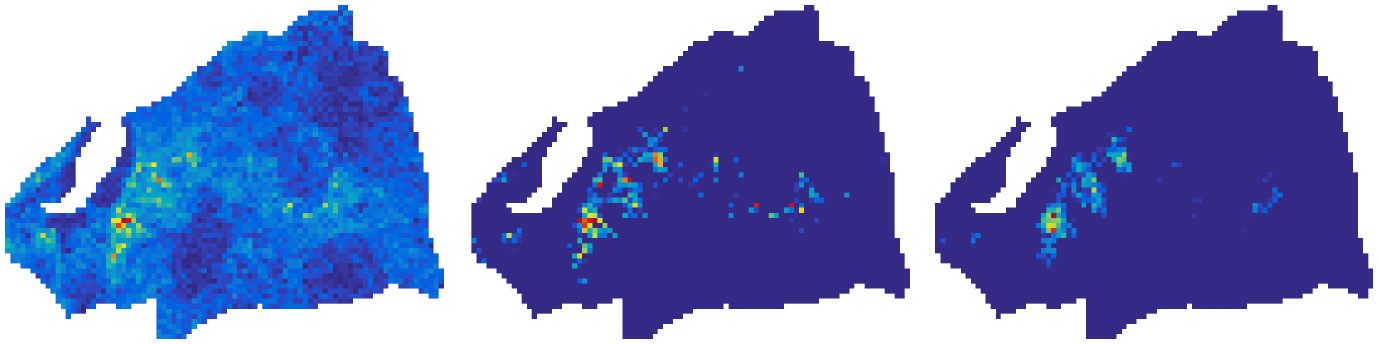} };
		\node[inner sep=0pt] at (\absleft,0.8) (dummy) {\fosize \textbf{(A)} $m/z$ 788};
		\node[inner sep=0pt,yshift = -0.15cm,xshift=-2.7cm] at (A.south) {\fosize raw};
		\node[inner sep=0pt,yshift = -0.15cm,xshift=-0.1cm] at (A.south) {\fosize basic};
		\node[inner sep=0pt,yshift = -0.15cm,xshift=2.5cm] at (A.south) {\fosize average};
		\node[inner sep=0pt] at (0,-1.25) {};
	\end{tikzpicture}
		
	\begin{tikzpicture}
		\node[inner sep=0pt](A) { \includegraphics[width=0.9\textwidth]{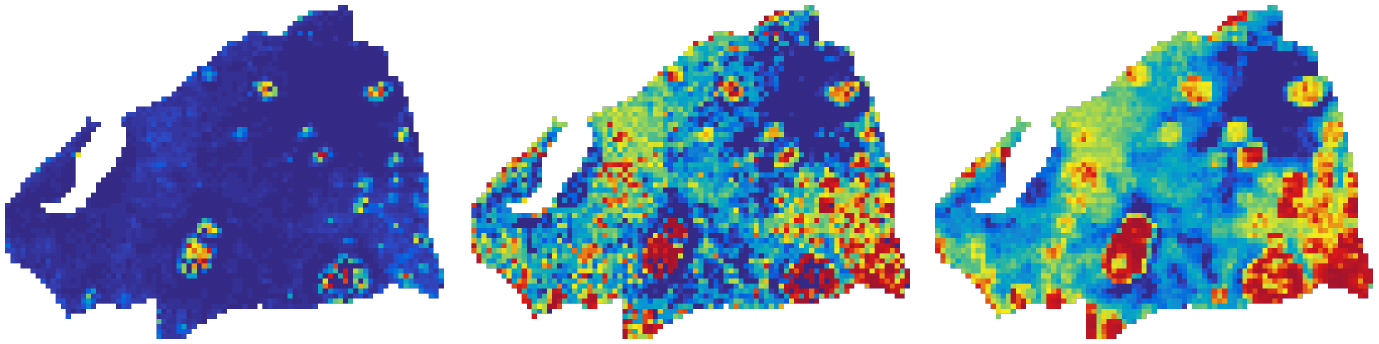} };
		\node[inner sep=0pt] at (0,-1.25) {};
		\node[inner sep=0pt] at (\absleft,0.8) (dummy) {\fosize \textbf{(B)} $m/z$ 843};
		\node[inner sep=0pt,yshift = -0.15cm,xshift=-2.7cm] at (A.south) {\fosize raw};
		\node[inner sep=0pt,yshift = -0.15cm,xshift=-0.1cm] at (A.south) {\fosize basic};
		\node[inner sep=0pt,yshift = -0.15cm,xshift=2.5cm] at (A.south) {\fosize average};
	\end{tikzpicture}

	\begin{tikzpicture}
	
		\node[inner sep=0pt](A) { \includegraphics[width=0.9\textwidth]{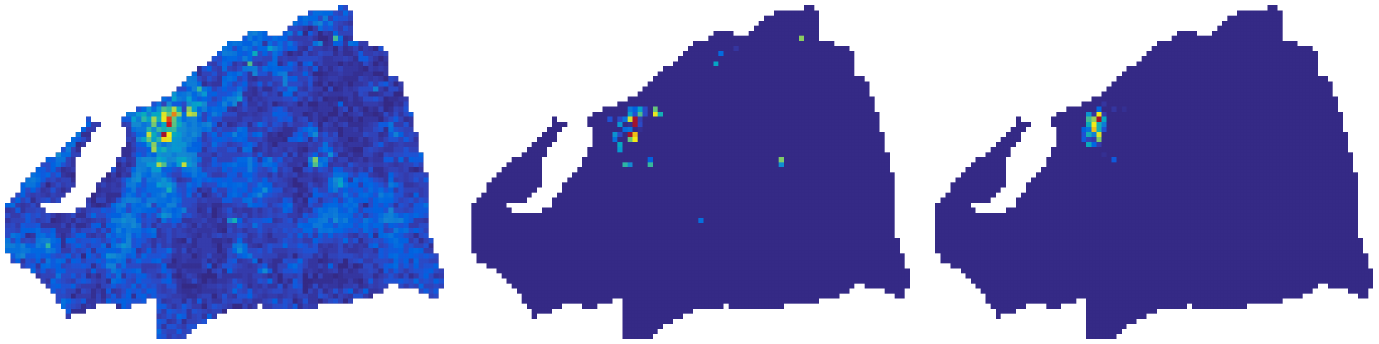} };
		\node[inner sep=0pt] at (0,-1.25) {};
		\node[inner sep=0pt] at (\absleft,0.8) (dummy) {\fosize \textbf{(B)} $m/z$ 942};
		\node[inner sep=0pt,yshift = -0.15cm,xshift=-2.7cm] at (A.south) {\fosize raw};
		\node[inner sep=0pt,yshift = -0.15cm,xshift=-0.1cm] at (A.south) {\fosize basic};
		\node[inner sep=0pt,yshift = -0.15cm,xshift=2.5cm] at (A.south) {\fosize average};
	\end{tikzpicture}

	\begin{tikzpicture}
	
		\node[inner sep=0pt](A) { \includegraphics[width=0.9\textwidth]{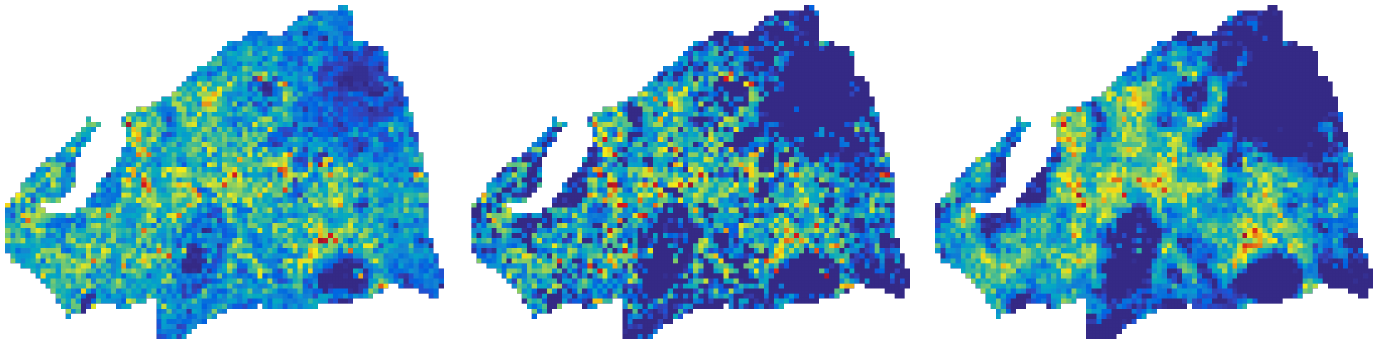} };
		\node[inner sep=0pt,yshift = -0.15cm,xshift=-2.7cm] at (A.south) {\fosize raw};
		\node[inner sep=0pt,yshift = -0.15cm,xshift=-0.1cm] at (A.south) {\fosize basic};
		\node[inner sep=0pt,yshift = -0.15cm,xshift=2.5cm] at (A.south) {\fosize average};
		\node[inner sep=0pt] at (0,-1.3) {};
		\node[inner sep=0pt] at (\absleft,0.8) (dummy) {\fosize \textbf{(D)} $m/z$ 1236};
	\end{tikzpicture}

	\caption{Selected $m/z$ images}
	\label{fig:2C}
\end{subfigure}
\end{tabular}

\caption{Results for the lung tissue data set: (a) mean spectra of the original data (top) and after peak picking (bottom), (b) an example spectrum showing the detection of isotope patterns by using long or short windows in the Gabor frame and (c) selected $m/z$ images of the raw data (\textit{raw}), after peak picking without spatial information (\textit{basic}) and including spatial information based on a $3\times3$ average filter (\textit{average}). Triangles indicate the selected $m/z$ values in the mean spectrum.}
\label{fig:lung}
\end{figure*}
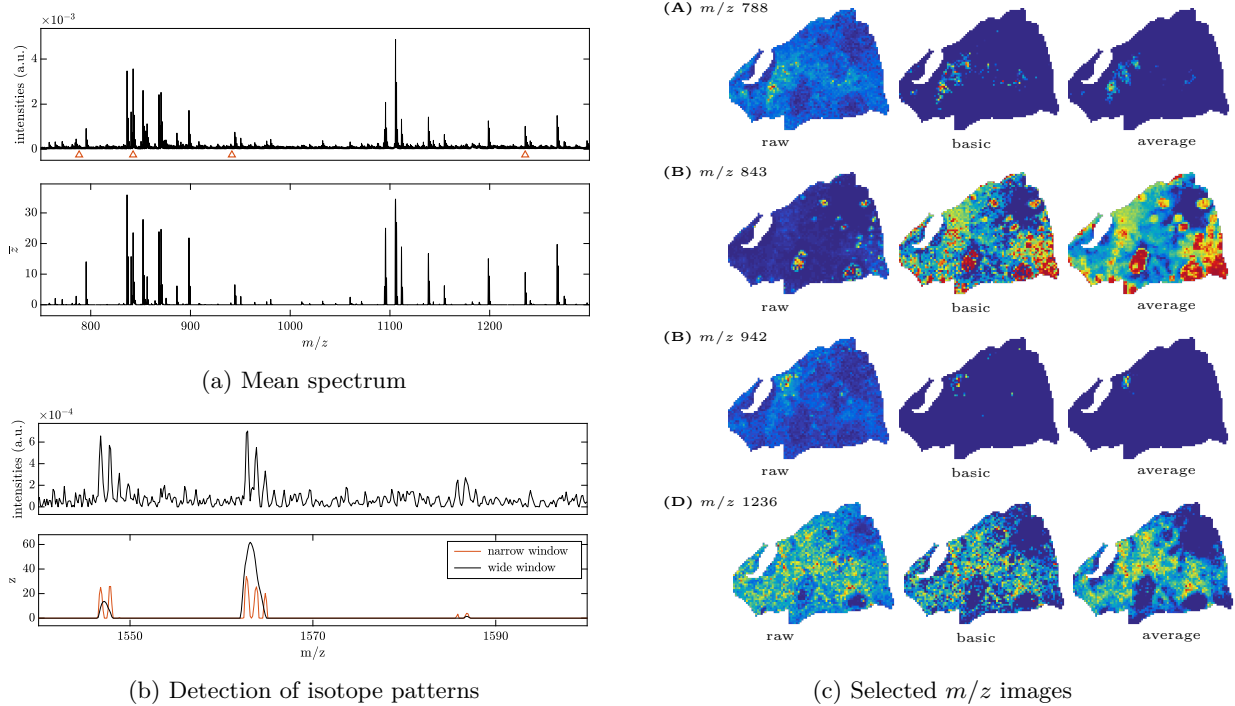

The mean spectrum of the resulting denoised data $z$ retains only 15\% all 20992 $m/z$ images with non-zero information. A section of the resulting mean spectrum is depicted in Figure \ref{fig:2A}, showing the similarity of the data set after peak picking with the original one.  Again, note that peak intensities of both spectra are not correlated any more. Nonetheless, considering $m/z$ images based on denoised data may reveal structures which would remain hidden in the original data set.   

In order to show this, $m/z$ images after \textit{basic} and \textit{average} peak picking are visualized in Figure \ref{fig:2C}.
Corresponding mass-to-charge values are indicated in the mean spectrum in Fig.~\ref{fig:2A} by triangles.
Figures (A) and (C) in Fig.~\ref{fig:2C} show $m/z$ images corresponding to small intensities in the mean spectrum, which are sparsely localized. On the other hand, Figures (B) and (D) in Fig.~\ref{fig:2C} reveal certain structures which are not readily visible in the original data, even with hotspot removal applied. So-called hotspots are removed by setting a certain percentage of the largest peak intensities to the lowest intensity among them. In particular, the band structure of localized dots on the left hand side of the lung tissue in Fig.~\ref{fig:2C} (B) becomes visible only when using the spatially aware approach.

With the increased mass resolution of the reflector mode TOF isotope patterns are well separated. The proposed approach based on Gabor frames with short windows is capable of detecting each isotopic peak separately. This is crucial for accurate protein identification \cite{nicolardi2015}. On the other hand, wider window functions can be used to find entire isotope patterns without resolving isotopic distributions, which may help to detect isotopic envelopes \cite{XIAO201741}. 
Both approaches are illustrated in Fig.~\ref{fig:2B} by showing part of an original spectrum and the same spectrum after applying the peak picking algorithm with a narrow and a wide window. It can be seen that even the isotope pattern which is buried in noise ($m/z\approx 1586$) can be reliably detected.

\section{Conclusion}
A novel spatially aware peak picking algorithm for MALDI MSI data has been introduced. It is based on sparse estimations of frame multiplier masks, measuring similarities between overlapping parts of a spectrum.  A slight modification of the algorithm also allows for incorporating spatial information into the peak picking process. This combines the usual three preprocessing steps in Figure \ref{fig:maldipipeline} into a single step, reducing computational complexity and simplifying parameter choices. 

On simulated data the accuracy of the peak picking algorithm shows a significant increase, while at the same time reducing false discovery rates by more than 50\% compared to a state-of-the-art algorithm. The numerical results verify that baseline effects can be ignored when using wavelet frames. Furthermore, the proposed algorithm is applied to two real MALDI-TOF data sets highlighting the advantages of including spatial information in the peak picking process. Although the denoised data does not correspond to original intensities any more, its visualization has been shown to detect spatial patterns which might otherwise remain unnoticed. Hence, the algorithm can either be used as an actual peak picking approach indicating peak locations, or as a denoising approach exposing hidden peptide structures. The latter might be advantageous as a preprocessing step prior to segmentation or clustering algorithms \cite{Alexandrov2013a,Alexandrov2011}. 

As the sampling of the $m/z$ axis is not equidistant over the mass range, the peak width changes with increasing mass-to-charge ratio. To overcome this problem, the proposed peak picking approach can be utilized with Gabor frames where the window size increases with corresponding $m/z$ sampling distance. For wavelet frames, a similar behavior can be realized by adapting discrete scales of the frame to corresponding $m/z$ values. 
Estimating peak parameters such as peak area or peak width has not been addressed so far. In \cite{Wijetunge2015} it is stated, that the peak area is more important than actual peak intensity when estimating molecular abundances. As demonstrated in \cite{Zhang2015} these parameters can be easily extracted from corresponding wavelet coefficients. For Gabor frames such a quantization of peaks is still an open topic and leaves room for further improvement. 

In real data sets the number of peaks present is generally unknown, which makes a proper choice of the regularization parameter $\lambda$ quite challenging. The regularization parameters used for peak picking in the previous section have been chosen empirically based on the performance of the algorithm applied to the mean spectrum. This means, for example, that $\lambda$ can be chosen such that a certain number of peaks are detected, or such that a certain percentage of $m/z$ values in the mean spectrum contain peaks. Other possible choices are still subject of current research.

With increasing mass resolution, for example using MALDI-Fourier transform ion cyclotron resonance (FT-ICR), the reliability to discriminate metabolites significantly improves to the range of millidaltons \cite{Palmer2017}. However, FT-ICR spectroscopy with high spectral resolution also results in larger noise as well as a larger data set size of up to $500$\,GB \cite{buck2105}. The proposed algorithm is capable of reliably detecting spectral patterns in noisy data while at the same time reducing the size of large data sets, making it also a good approach for analyzing MALDI-FT-ICR data.

\section*{Appendix}
\subsection*{Proof of Theorem 1}
\begin{proof}
Obviously, $m=1$ whenever $\abs{c_1}$ is trivial. Assuming that $\abs{c_1}$ is non-trivial leads to
\begingroup
\addtolength{\jot}{0.1em} 
\begin{align}
 0 &\in \nabla_m\left(\frac{1}{2}\norm{\abs{c_2}- m \abs{c_1}}^2_2\right) + \lambda \partial \left(\norm{m - 1}_1\right) \nonumber\\
 \Leftrightarrow \quad 0 & \in  m \abs{c_1}^2 -\abs{c_2}\abs{c_1} + \lambda \partial \left(\norm{m - 1}_1\right) \nonumber\\
 \Leftrightarrow \hspace{7.5pt} m & \in \frac{\abs{c_2}}{\abs{c_1}} - \frac{\lambda}{\abs{c_1}^2} \partial \left(\norm{m - 1}_1\right). \label{eq:pr_mkl}
\end{align}
\endgroup
The subdifferential of the $\ell_1$-norm consists of the following subgradients, which can be evaluated for each coefficient separately
\begin{equation}
\partial \abs{m - 1} = \left\{ 
	\begin{array}{ll} 
		\{1\} & \mbox{if } m > 1 \\ 
		\{-1\} & \mbox{if } m < 1 \\
	  \left[-1,1\right] & \mbox{if } m = 1 \\
	\end{array}  
\right. .
\label{eq:subdiffm}
\end{equation}
Considering all three cases in \eqref{eq:pr_mkl} leads to the closed form solution 
\begin{equation}
m = \left\{
	\begin{array}{ll}
		\frac{\abs{c_2}\abs{c_1} - \lambda}{\abs{c_1}^2} & \mbox{if } \frac{\abs{c_2}}{\abs{c_1}} - 1 > \frac{\lambda}{\abs{c_1}^2} \\[8pt]
		\frac{\abs{c_2}\abs{c_1} + \lambda}{\abs{c_1}^2} & \mbox{if } \frac{\abs{c_2}}{\abs{c_1}} - 1 < -\frac{\lambda}{\abs{c_1}^2} \\[8pt]
		1 &\mbox{if } -\frac{\lambda}{\abs{c_1}^2} \leq \frac{\abs{c_2}}{\abs{c_1}} - 1 \leq \frac{\lambda}{\abs{c_1}^2}
	\end{array}
	\right..
\label{eq:clfosol1}
\end{equation}
With $y = \abs{c_2}\abs{c_1}^{-1}$, the equivalence of \eqref{eq:clfosol1} and Eq.~(9) in the main manuscript follows directly: assuming $1-\lambda\abs{c_1}^{-2}\abs{y-1}^{-1}\leq 0$ in (9) is equivalent to the third row in \eqref{eq:clfosol1}. On the other hand, $1-\lambda\abs{c_1}^{-2}\abs{y-1}^{-1} > 0$ leads to the first and second row of \eqref{eq:clfosol1} by considering the cases $y-1>0$ and $y-1<0$, respectively.
\end{proof}

\section*{Supplemental}
The source code of our algorithm can be found online under \url{https://github.com/flieb/MALDIPeakDetection}.

\section*{Acknowledgment}

The authors thank Jan H.~Kobarg (SCiLS, Bremen, Germany) for providing the rat brain data set and Janina Oetjen (then at University of Bremen, MALDI Imaging Lab, Bremen, Germany) for acquiring the lung tissue MALDI data. The FFPE lung tissue section has kindly been provided by Rita Casadonte (Proteopath, Trier, Germany).

\setstretch{1.1}
\normalsize
\bibliographystyle{plainnat}  
\bibliography{refs}

\end{document}